\documentclass[aps,prc,twocolumn,floatfix,nofootinbib,preprintnumbers,superscriptaddress,amsmath,amssymb]{revtex4-1}

\usepackage{graphicx}
\usepackage{dcolumn}
\usepackage{bm}
\usepackage{hyperref}
\usepackage[mathlines]{lineno}

\begin{document}

\title{Optimizing multilayer Bayesian neural networks for evaluation of fission yields}

\author{Zi-Ao Wang}
\affiliation{State Key Laboratory of Nuclear
Physics and Technology, School of Physics, Peking University,  Beijing 100871, China}

\author{Junchen Pei}\email{peij@pku.edu.cn}
\affiliation{State Key Laboratory of Nuclear
Physics and Technology, School of Physics, Peking University,  Beijing 100871, China}

\begin{abstract}

The Bayesian machine learning is a promising tool for the evaluation of nuclear fission data but its potential capability has not been fully realized.
We attempt to optimize the performances of the multilayer Bayesian neural networks for evaluations of fission yields.
The influences of adjustments of learning data, activation functions, network structures have been studied.
In particular, negative values of net functions have been penalized to avoid non-physical inferences of fission yields.
Presently the network with double hidden layers has optimal performances compared to the single-layer or deeper networks.
These studies are essential for further developments of  precise evaluation methods.

\end{abstract}

\maketitle


\section{Introduction}

Nuclear data evaluation is a crucial process that connects nuclear data, nuclear modeling and nuclear applications.
In particular, nuclear fission data is the  key ingredient in many nuclear applications~\cite{nd}.
For example, high-precision and reliable neutron-induced fission product yield (FPY) distributions of actinides are very valuable.
However,
experimental measurements of FPY with continuous incident neutron energies is extremely difficult and insufficient.
In major nuclear data libraries (ENDF~\cite{endf}, JENDL~\cite{jendl}, JEFF~\cite{jeff}, CENDL~\cite{cendl}, etc.),  complete evaluations of FPY are only available for neutron incident energies
around thermal energies, 0.5 MeV and 14 MeV.
There are some incomplete experimental FPY at other energies with large uncertainties.
On the other hand,  theoretical descriptions of fission observables are very challenging~\cite{future,schunck,gef1,qiang}.
Microscopic fission models are promising but are not ready yet for accurate quantitative applications.
Therefore, the prediction and evaluation of energy dependent  FPY for fast reactors
are very anticipated.

Machine learning is a very powerful tool for learning and inference from complex big data.
In recent years, machine learning has attracted great interests in various physics disciplines.
Recently, it was shown that Bayesian neural network (BNN) can be used for evaluations of
incomplete
fission yields with uncertainty quantifications~\cite{fissionPRL2019}.
The machine learning has been used in nuclear physics with increasing interests~\cite{ai}, such as  the extrapolation of nuclear masses~\cite{Utama2016PRC,Niu2018PLB,Neufcourt2018PRC},
 fission yields~\cite{fissionPRL2019,amy},  various nuclear structures~\cite{Niu2019PRC,bai,keeble,Lasseri,Utama2016JPG,jiang} and reaction observables~\cite{Ma2020CPC014104,Ma2020CPC124107,amy2}.
The machine learning has also been widely applied in other physics subjects, such as the constrains of equation of state of neutron stars from gravitational wave signals~\cite{gw}
and for facilitating the lattice QCD calculations~\cite{qcd}.
Conventionally, the evaluation of fission yields is based on the least-squares adjustments of parameters of various phenomenological models~\cite{england}, such as
the Brosa model and the GEF model~\cite{brosa,gef}. These evaluations could not be applicable when very few experimental data points are available.
Machine learning is promising for developing new evaluation methods of nuclear data, in regarding to handle various correlated fission observables with large discrepancies and uncertainties.

Previously we have demonstrated that BNN can be used for evaluation of incomplete fission mass yields~\cite{fissionPRL2019} and fission charge yields~\cite{qiao}.
In this work, we aim to improve the performances of multilayer Bayesian neural networks for precise evaluation of fission yields.
The potential capability of BNN evaluation has not been fully realized.
In principle, the BNN approach can be optimized for specific applications.
For example, the learning performance and overfitting are two competing issues in machine learning.
Therefore, more studies about the sensitivity of configurations of neural networks are needed.
In this work, we plan to study the influences of adjustments of learning data, the choice of activation functions,
the structure of multilayer networks, and the penalty of negative values to constrain the fission yields.
Finally, as an example,  we demonstrate that BNN is used for precise evaluations of fission mass yields of $^{239}$U.

\section{The theoretical framework}

The BNN approach~\cite{Neal1996} to statistical regression inference is based on Bayes' theorem, which provides a connection between a given hypothesis(in terms of  problem-specific beliefs for a set of parameters) and a set of data  to a posterior probability  that is used to make predictions on new inputs.
The BNN approach adopts probability distributions as connection weights and is naturally suitable for uncertainty quantifications,
in contrast to standard neural networks which optimize definite values for connection weights.
The basic BNN is written as,
\begin{equation}
    p(\omega |{\rm x},t)=\frac{p({\rm x}, t| \omega)p(\omega)}{p({\rm x},t)},
                                                                  \label{eqn.01}
\end{equation}
where \emph{p}(x,t$\mid$$\omega$) is the `likelihood' that a given model describes the data and \emph{p}($\omega$) is the prior distribution of the parameters $\omega$; \emph{x} and \emph{t} are input and output data; \emph{p}($\omega$$\mid$x, t) is the the posterior distribution, i.e., the probability distribution of parameters $\omega$ after considering the data (x, $t$); \emph{p}(x, $t$) is a normalization factor which ensures the integral of posterior distribution is one.

We adopt a Gaussian distribution for the likelihood based on a cost function, which is written as
\begin{equation}
    p({\rm x},t\mid \omega)=\exp(-\chi ^{2}/2),
                                                                  \label{eqn.02}
\end{equation}
where the cost function $\chi$$^{2}$($\omega$) reads:
\begin{equation}
    \chi^{2}(\omega)=\sum^{N}_{i=1}(\frac{t_{i}-f({\rm x_{i}},\omega)}{\Delta t_{i}})^{2},
                                                                  \label{eqn.03}
\end{equation}
Here \emph{N} is the number of  data points, and $\Delta$$t_{i}$ is the associated noise scale which is related to specific observables. The net function \emph{f}(x$_{i}$, $\omega$) depends on the input data x$_{i}$ and the model parameters $\omega$. In this work, the inputs of the network are given by x$_{i}$=\{A$_{fi}$, Z$_{i}$, A$_{i}$, E$_{i}$\}, which include the mass number A$_{fi}$ of the fission fragments, the charge number Z$_{i}$ and mass number A$_{i}$ of the fission nuclei and the excitation energy of the compound nucleus E$_{i}$=e$_{i}$+S$_{i}$(e$_{i}$ and S$_{i}$ are incident neutron energy and neutron separation energy, respectively); $t_{i}$ are the fission mass yields. For the evaluation of fission yields, it is crucial to learn the yields individually and it is difficult to learn complete distributions as a target.  Fortunately, the normalization of fission yields is kept within an uncertainty of 2$\%$.

The posterior distributions are obtained by learning the given data. With new data $\rm {x_n}$,
we make predictions by integrating the neural network over the posterior probability density of the network parameters $\omega$,
\begin{equation}
    \langle f_{n} \rangle=\int f({\rm x_{n}},\omega)p(\omega \mid{\rm x},t)d\omega,
                                                                  \label{eqn04}
\end{equation}
The high-dimensional integral in Eq.\ref{eqn04} is approximated by Monte Carlo integration in which the posterior probability \emph{p}($\omega$$\mid$x, $t$) is sampled using the Markov Chain Monte Carlo method~\cite{Neal1996}.

In BNN we need to specify the form of the functions \emph{f}(x, $\omega$) and \emph{p}($\omega$). In this work, we use a feed-forward neural network model defined the function \emph{f}(x, $\omega$). That is
\begin{equation}
    f({\rm x},\omega)=a+\sum^{H}_{j=1}b_{j}\tanh(c_{j}+\sum^{I}_{i=1}d_{ji}{\rm x}_{i}),
                                                                  \label{eqn.05}
\end{equation}
where \emph{H} is the number of neurons in the hidden layer, \emph{I} denotes the number of input variables and $\omega$=\{$a$, $b_{j}$, $c_{j}$, $d_{ji}$\} is the model parameters, $a$ is bias of output layers, $b_{j}$ are the weights of output layers, $c_{j}$ is bias of hidden layers, and $d_{ji}$ are weights of hidden layers. In total, the number of parameters in this neural network is 1+(2+\emph{I})$\times$\emph{H}.
We adopt the commonly used tanh as the activation function and other non-linear activation functions have also been tested in this work.
The confidential interval (CI) at 95\% level is given for uncertainty quantifications in this work.
More details about  BNN  can be found in Ref.~\cite{Neal1996}.

\begin{figure}[t]
\centering
\includegraphics[width=0.42\textwidth]{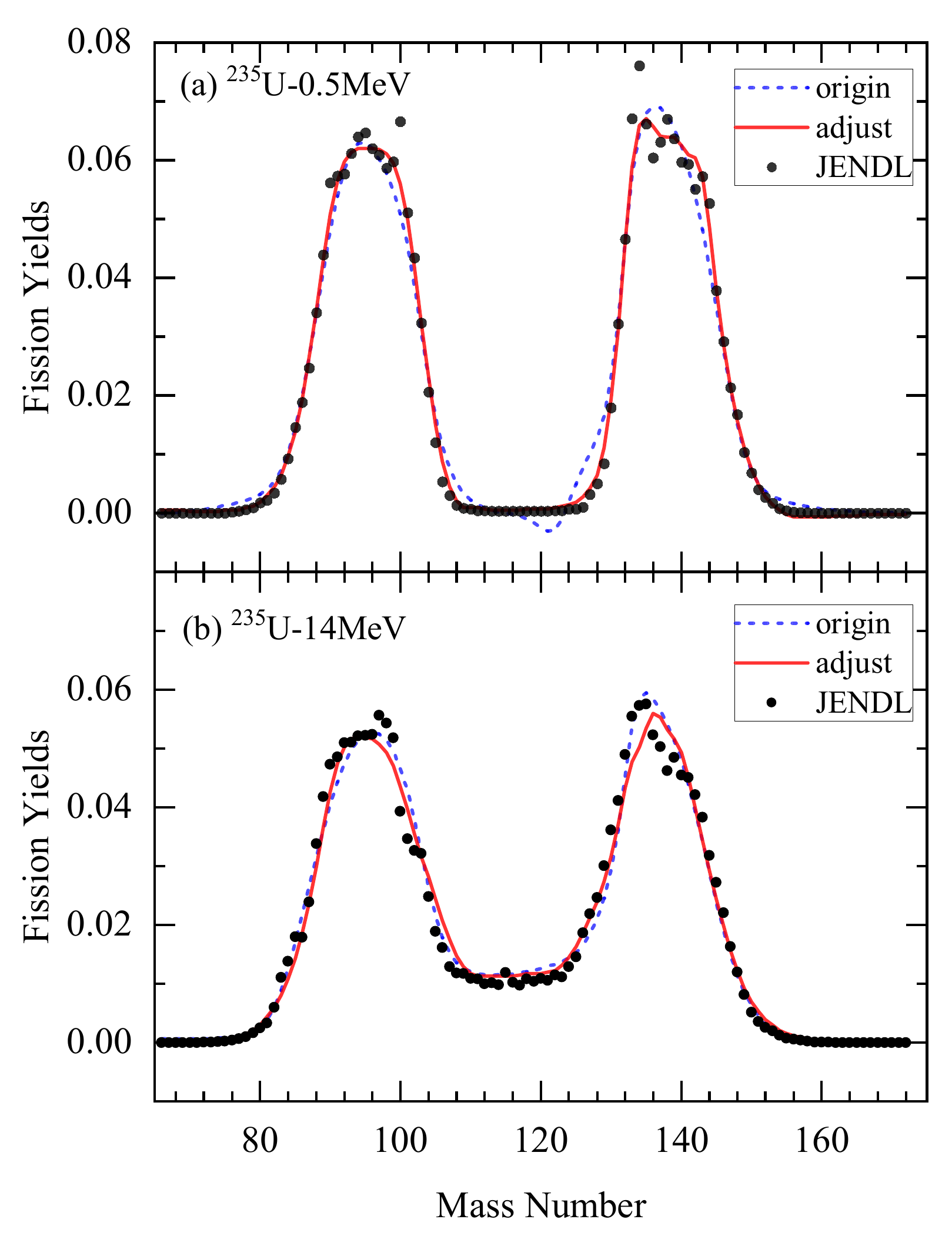}
\caption{
The BNN learning performances are shown for neutron induced fission yields of $^{235}$U at 0.5 and 14 MeV respectively.
The results obtained by learning the original fission data or the adjusted data are compared. The adjustment of data denotes
a linear transformation so that the ranges of data values are within ($-$0.9, 0.90). The learning target is taken from the evaluated data of JENDL~\cite{jendl}.
\label{FIG1}
}
\end{figure}

\begin{figure}[t]
\centering
\includegraphics[width=0.42\textwidth]{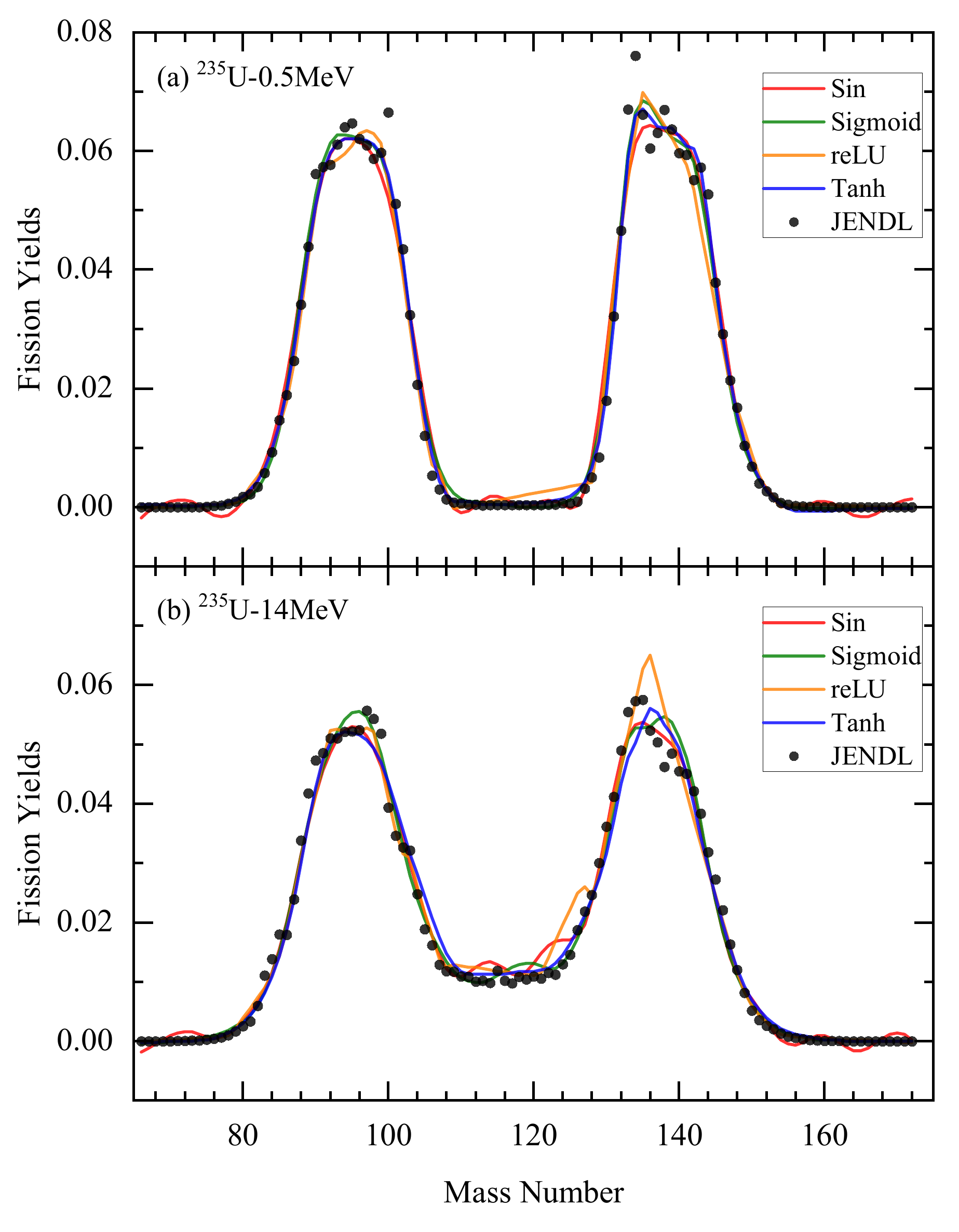}
\caption{
The learning performances with different activation functions (sine, tanh, sigmoid, ReLU) are shown for fission yields of $n+^{235}$U at 0.5 and 14 MeV respectively.
\label{FIG2}
}
\end{figure}

\begin{figure}[t]
\centering
\includegraphics[width=0.42\textwidth]{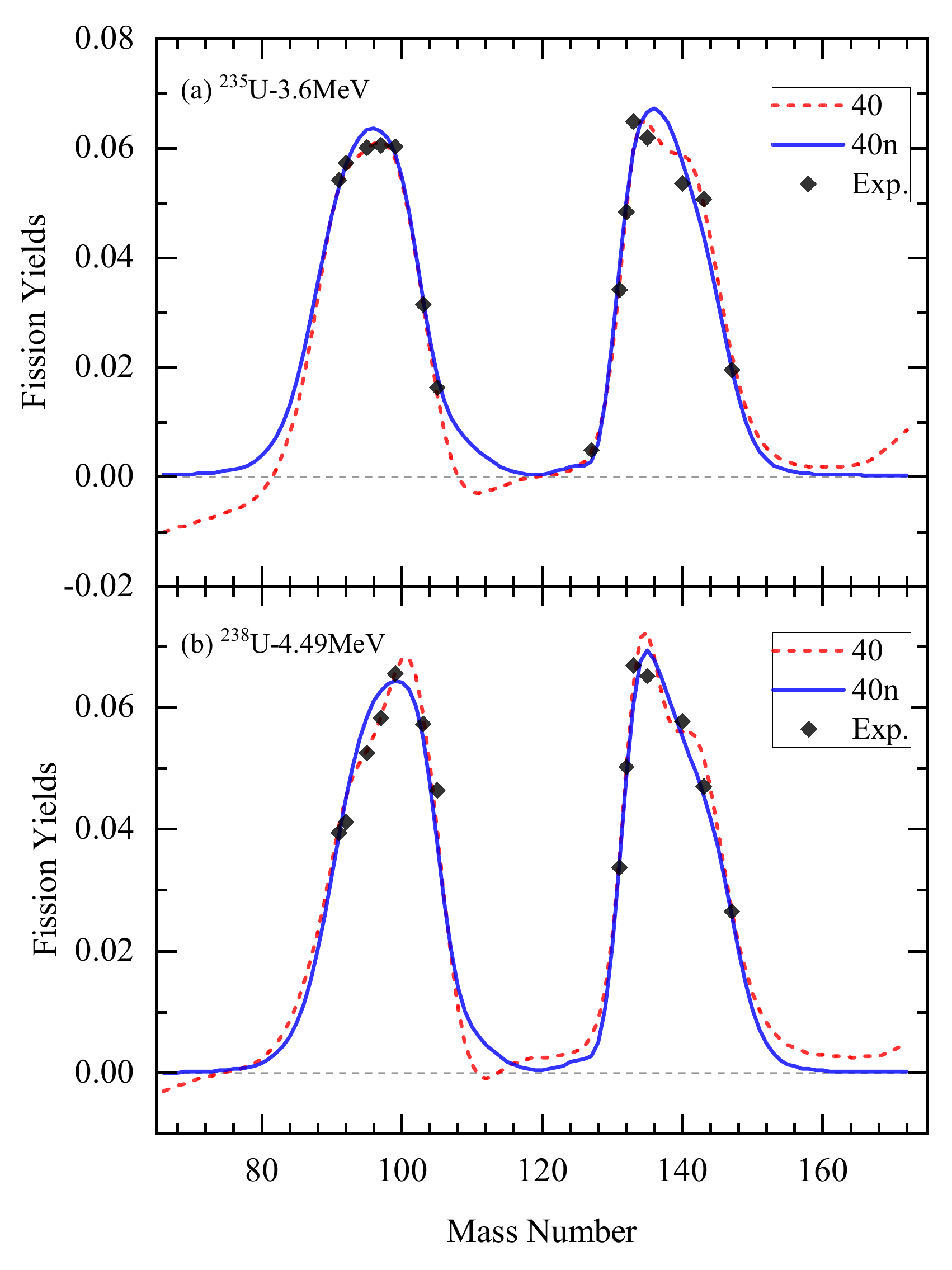}
\caption{
The BNN evaluation of incomplete neutron induced fission yields of $^{235}$U at 3.6 MeV (a), and $^{238}$U at 4.49 MeV (b).
The normal BNN evaluation used one hidden layer with 40 neurons. The BNN results with learning penalty on negative values are shown for comparison, denoted as `40n'.
The experimental data is taken from~\cite{expt2016}.
\label{FIG3}
}
\end{figure}

\section{The results and discussions}

\begin{figure}[t]
\centering
\includegraphics[width=0.48\textwidth]{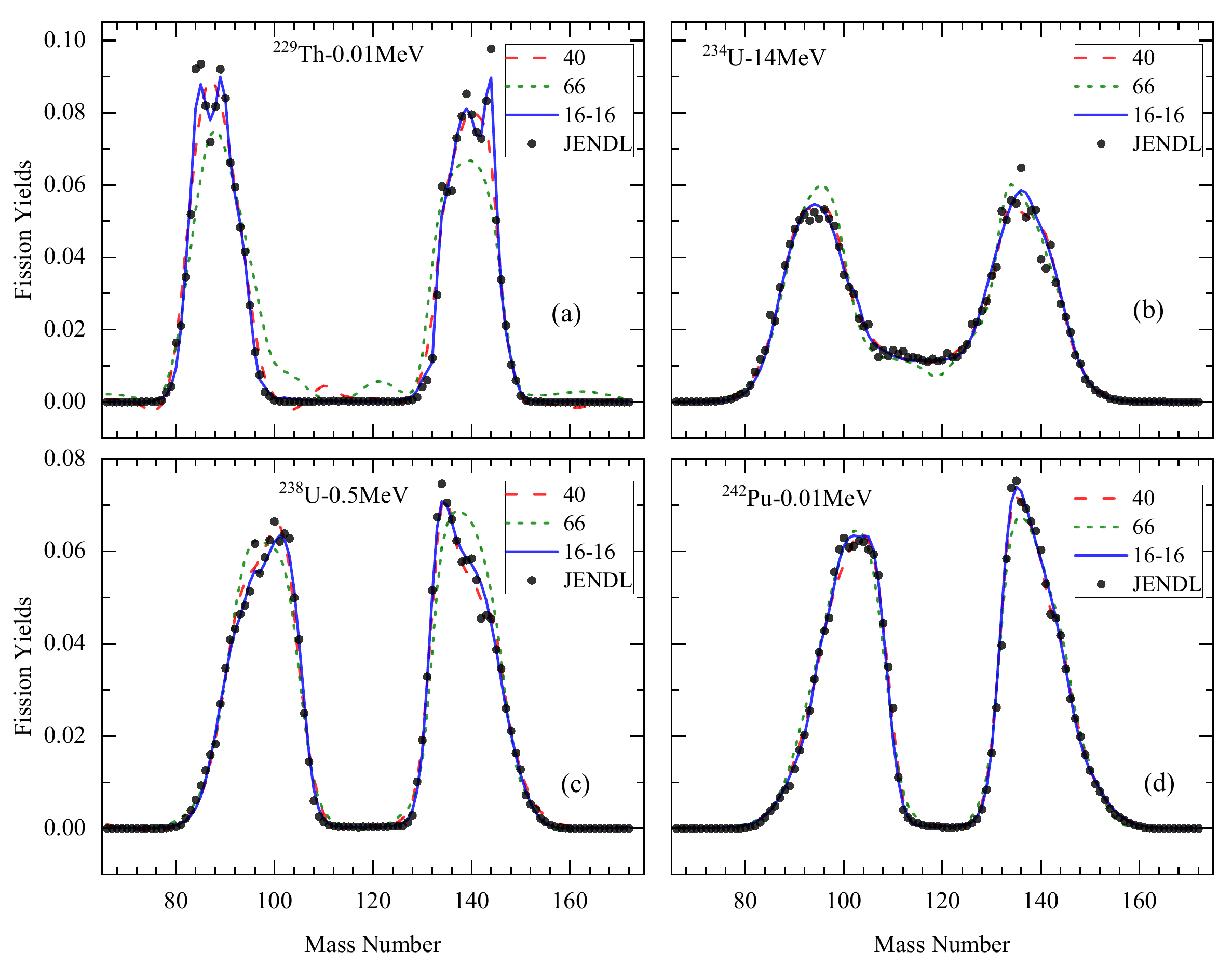}
\caption{
BNN learning performances are shown with different networks: one layer with 40 neutrons, one layer with 66 neurons,
two layer with 16-16 neutrons.
\label{FIG4}
}
\end{figure}

\begin{figure}[t]
\centering
\includegraphics[width=0.45\textwidth]{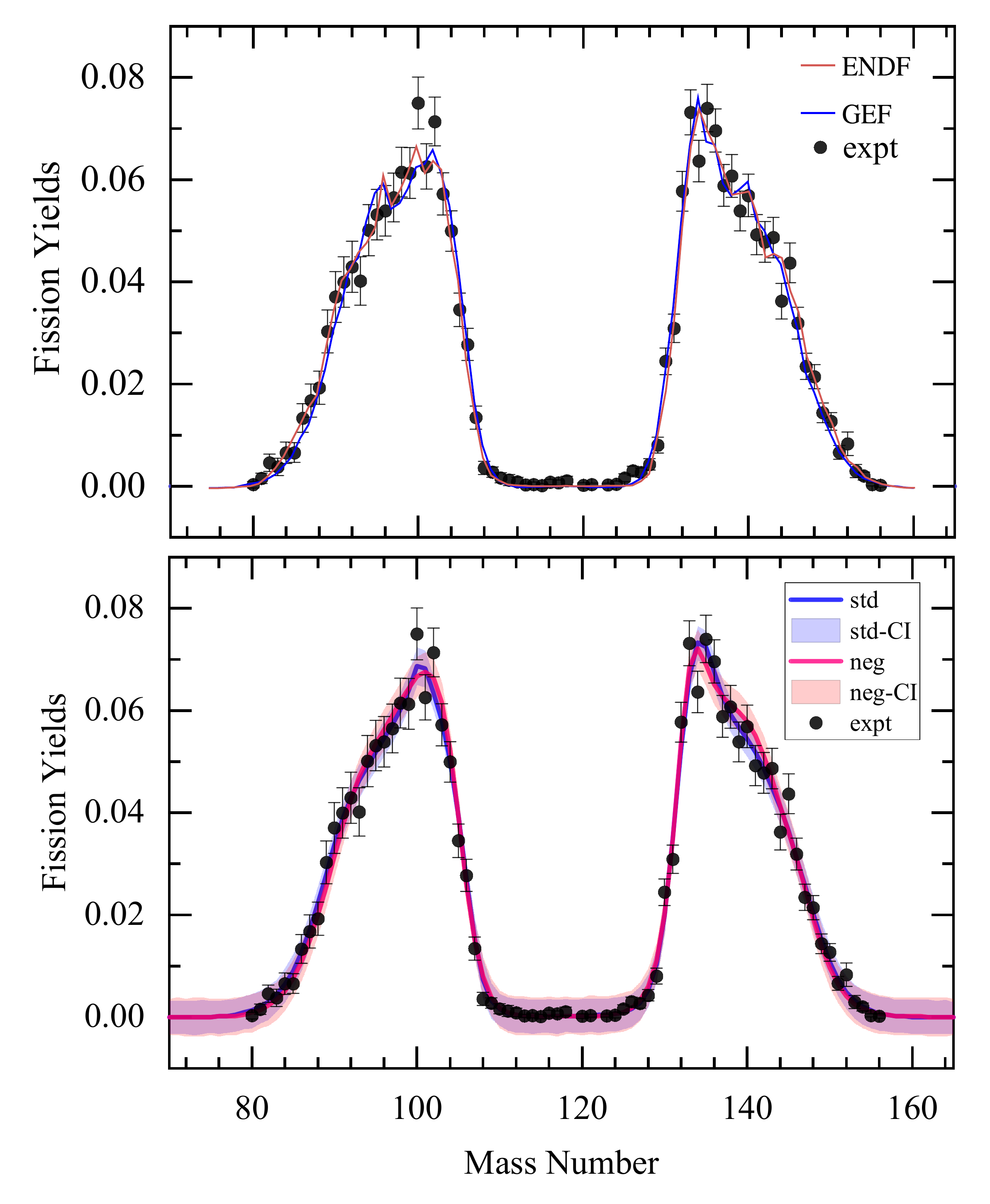}
\caption{
The evaluations of fission yields of compound nucleus $^{239}$U are shown. The experimental data is taken from~\cite{expt239}.  GEF and ENDF evaluations are shown in (a), which are taken from~\cite{expt239}.
The double-layer BNN evaluations without (denoted as `std') and with penalty on negative values (denoted as `neg') are shown in (b). 
The shadows in (b) denote the BNN confidential interval at 95$\%$.
\label{FIG5}
}
\end{figure}

Firstly we study the influences of adjustments of the input data.
In principle, the network can take any data without range limits.
However, we speculate that the network is more sensitive to a specific range of data. 
For example, the tanh activation function is not sensitive to large values and the
response is within the saturation range.
Actually, the tanh function is more sensitive from $-$0.9 to 0.9. 
To this end, we performed  linear transformations of the original data so that
both input and output data are within the range of ($-$0.9, 0.9).
Present calculations employed 5350 evaluated fission yield data of 30 nuclei from JENDL, which is similar to
our previous work~\cite{fissionPRL2019}. 
Fig.\ref{FIG1} displays the learning results of $n+^{235}$U at 0.5 MeV and 14 MeV respectively. 
Note that calculations in Fig.(1, 2, 3) employed one hidden layer with 40 neurons. 
It can be seen that the learning performance  with the original data is not satisfactory at 0.5 MeV.
On the other hand, the learning results with adjusted data are rather good without unphysical oscillations. 
At 14 MeV, the two results are comparable. 
Previous studies also show that machine learning is difficult to describe the fission yields at low excitation energies with more quantum effects~\cite{qiao}. 
This points out that the adjustment of ranges of input data indeed has advantages for BNN
performances. 

In neural networks, it is crucial to use nonlinear activation functions to compute nontrivial problems. 
There are various activation functions been adopted in neural networks~\cite{activation} and it is interesting to 
choose a particular activation function for specific problems. 
 In this work, we did testing calculations with
 tanh function, sigmoid function, ReLU (Rectified Linear Unit) function, and sine function, as shown in Fig.\ref{FIG2}.
We see that all activation functions can largely reproduce the fission yields.
However, sine and ReLU functions result in some unphysical oscillations. 
The results of tanh are slightly better than that of sigmoid.
With 5350 points, the  total $\chi_N^2$=$\sum\limits_i (t_i-f(x_i))^2/N$ for tahn, sigmoid, ReLU and sine activation functions
are  5.78$\times$10$^{-6}$, 6.37$\times$10$^{-6}$, 8.79$\times$10$^{-6}$
and 7.56$\times$10$^{-6}$, respectively. 
Note that the total learning steps are 10$^5$ for all activation functions. 
We also see that sine and tanh functions can have unphysical negative fission yields, which are suppressed in sigmoid and ReLU functions. 
Generally, we see that tanh activation function has the best learning performance.

In evaluation of fission yields, the physical output values should always be positive. 
However, the output ranges of tanh function could be negative. 
This is a serious problem for inferences when data is sparse~\cite{fissionPRL2019}. 
To solve this problem, we add penalty to constrain the fission yields.
Actually, the weights of the likelihood function are increased when negative values
appear. 
Fig.\ref{FIG3} shows that evaluation of incomplete fission yields from neutron included fission
of $^{235}$U at 3.6 MeV and $^{238}$U at 4.49 MeV. 
We see that BNN evaluation without penalty leads to
serious unphysical negative values. 
The results obtained with penalized learning have much better performance when
fission yields are close to zero.
As a compromise, we also see that the data points can be better reproduced by BNN without penalty.
Nevertheless, the penalized learning is necessary to avoid the unphysical negative values. 
This is also a successful attempt to build physical constraints into machine learning, towards
a physics-guided machine learning.

Next we explored the optimal structure of neural networks for evaluation of fission yields. 
Fig.\ref{FIG4} shows the learning results of one hidden layer with 40 neurons and  66 neurons, and two hidden layers with 16-16 neurons.
We can see that results of one layer with 40 neurons are largely satisfactory.
The double layers with 16-16 neurons has the best performance. 
The learning performance of one layer with 66 neurons are not so good with some oscillations. 
The total $\chi_N^2$ of 40, 66 and 16-16 networks are 5.78$\times$10$^{-6}$, 4.35$\times$10$^{-6}$ and
3.43$\times$10$^{-6}$, respectively. 
The average performance of one layer of 66 neurons is better than that of 40 neurons.
But the performance of 66 neurons are not always better than that of 40 neurons. 
Note that the number of connection weights of the 16-16 structure is close to that of the one layer of 66 neurons,
but the performance of the double layer structure is much better. 
This shows the limitation of the one hidden layer structure for complex data. 

Generally, with similar number of connection weights, the shallow network would be more dependent on the prior input.
In contrast, the deep network would be more dependent on its deduction capability. 
Therefore, for specific problems, there should be a balanced choice of network structure. 
We also tested network structures of 11-12-12, 9-10-10-10, 9-9-8-8-9, 8-8-8-8-7-7, 7-7-7-7-7-7-8 neurons
for 3, 4, 5, 6, 7 hidden layers, respectively. 
Note that all these structures have similar number of connection weights to that of the 16-16 structure. 
Correspondingly, the total $\chi_N^2$ are 4.07$\times$10$^{-6}$ (3 layers), 4.99$\times$10$^{-6}$ (4 layers), 4.64$\times$10$^{-6}$ (5 layers),
5.05$\times$10$^{-6}$ (6 layers), 4.94$\times$10$^{-6}$ (7 layers).
We see that deep networks have no advantages in this work. 
The best network is the double-layer structure of 16-16 neurons. 
In addition, the deep networks take much longer computing time to get convergence. 

Finally, we performed BNN evaluations of fission mass yields of $^{239}$U with the double layer network,
as shown in Fig.\ref{FIG5}.
The incomplete experimental data is taken from Ref.\cite{expt239}, in which the compound nucleus $^{239}$U is produced
through one neutron transfer reaction of $^{238}$U+$^{9}$Be. 
The recent experiments can obtain precise isotopic identification of some fragment isotopes. 
For comparison, BNN evaluations with and without negative penalty are given. 
We see that both evaluations are satisfactory with some slight discrepancies. 
It has to be pointed out that in the first peak, both evaluations underestimate the fission yields, indicating
that experimental yields at the first peak could be overestimated. 
In this respect, our evaluations are consistent with the GEF evaluation.
For the second peak, the largest value is at $A$=134 in BNN with negative penalty, which is consistent with
GEF and ENDF evaluations. 
This indicates that the fission yield at $A$=134 is underestimated by the experiment.
The fission yields at $A$=140 with negative penalty is slightly larger than that of the standard BNN evaluation, which is
consistent with GEF and ENDF evaluations. 
Generally, the shape of the second peak obtained with negative penalty is better consistent with GEF and ENDF evaluations,
compared to the standard BNN evaluation. 
We demonstrated that the BNN evaluation with the negative penalized function is essential to describe detailed peak structures
and obtain high precision evaluations.
The confidential interval from BNN with negative penalty is slightly larger than that from standard BNN  due to additional constraints. 
Note that the uncertainty propagation and quantification are very important in evaluations~\cite{talou}, which will be studied in detail in a forthcoming work.

\section{Summary}

In summary, we studied the multilayer Bayesian neural networks to improve its  performance for evaluations of fission yields. 
We investigated the influences of adjustments of ranges of input and output data for neural networks. 
It is useful to use a linear transformation to prepare the learning data set within a proper range so
that the response saturation range of the active function can be avoided. 
We also studied various active functions and found that tanh function has the best learning performance. 
In the evaluation of fission yields, there is a serious problem that unphysical negative values can appear in outputs of neural networks. 
We deployed the penalized function in likelihood function to constrain the outputs.
Thus the negative values can be much suppressed. This is a successful attempt to implement
physics constraints in neural networks. 
We also studied the various network structures from a single hidden layer to seven hidden layers.
We found that the double layer network is optimal for the present work. 
Finally we performed the BNN evaluation of fission yields of $^{239}$U.
The results show that BNN with negative penalty is essential to obtain detailed
peak structures and high precision evaluations. 
This work demonstrated that BNN is a promising tool for high precision evaluations of fission data
and work in this direction is in progress.

\begin{acknowledgments}
 This work was supported by  the
 National Key R$\&$D Program of China (Contract No. 2018YFA0404403),
  the National Natural Science Foundation of China under Grants No. 11975032, 11835001, 11790325, 11961141003.
\end{acknowledgments}

\nocite{*}


\end{document}